# Revisiting superradiance dynamics from single diamond nanocrystals with a physically consistent model for fluorescence decay

Jakub J. Borkowski, Artur Czerwinski & Piotr Kolenderski

Superradiance in nanodiamond samples at room temperature has attracted significant attention due to its potential applications in quantum technologies. The paper by C. Bradac et al.[1], supplemented with additional material[1], presents a theoretical model with numerical simulations that claim to reproduce the observed experimental behavior. Our analysis reveals that this model contains equations that produce unphysical results and do not accurately describe the experimental data. We propose corrected equations that yield physically meaningful predictions, highlighting the need to revise the theoretical framework for interpreting these experiments.

The structure of this work is outlined as follows. We begin by providing a concise review of the model proposed in ref. 1. Here, we present the principal equations along with explanations of the associated symbols. Subsequently, we present the examination of a scenario involving seven color centers, which is identical to the case explored in ref. 1. By employing computational software to implement the model[2], we obtain results that diverge from those presented in the aforementioned study. While we specifically discuss one case, other examples of discrepancies are detailed in the Supplementary Information. Furthermore, we propose equations that offer an accurate mathematical representation of the phenomenon under consideration. Through our proposed equations, we demonstrate that the instances previously misrepresented according to the model from ref. 1, now exhibit all required physical properties. Additionally, the fluorescence plots generated from our model show characteristic emission patterns. Complementing our study is a Supplementary Information, where we provide comments on individual corrected elements of the equations and elaborate on their proposed interpretations. Notably, the issues within the model from ref. 1 appear to be of various kinds; some may stem from computational inaccuracies, while others may originate from flawed physical premises.

For clarity and consistency, we introduce a convention to distinguish between models. The formalism outlined in the supplementary material to ref. 1 is referred to as **Model A**, while we designate the framework introduced in the present paper as **Model B**. We adhere to this distinction throughout the remainder of this text.

## Model A

The research discussed in ref. 1, as well as in our laboratory, involves diamond samples containing lattice defects where a nitrogen atom and a vacancy replace a pair of carbon atoms. These centers are subsequently excited by light, and the resulting spontaneous emission is observed. To facilitate a comparative analysis, we adopt identical labeling and experimental approaches as outlined in the referenced work[1]. Specifically, we divide the diamond into distinct domains denoted by $\sigma$, each containing a certain number of color centers $N_\sigma$, contributing to the total count $N$.

The foundational equation in constructing Model A is the Gorini-Kossakowski-Lindblad-Sudarshan (GKLS) master equation[3–6], which encompasses both non-phenomenological and phenomenological terms. These terms are expected to capture the intricacies of collective state defocusing and intersystem crossing phenomena within the context of superradiance. Each term in the equation is associated with constants $\gamma$, which are indicative of the characteristic rates governing these phenomena. These constants are devised analogously to the conventional decay rate, denoted as $\gamma^{(\sigma)}$, which typifies the spontaneous emission behavior of a single atom.

Institute of Physics, Faculty of Physics, Astronomy and Informatics, Nicolaus Copernicus University in Torun, Torun, Poland. e-mail: jborkowski@doktorant.umk.pl; aczerwin@umk.pl; kolenderski@umk.pl

Let us recapture the version of the GKLS equation with phenomenological factors following ref. 1:

$$\begin{aligned}\frac{d}{dt}P^{(\sigma)}_{J,M}(t)=&\,\gamma[(J(J+1)-M(M+1))P^{(\sigma)}_{J,M+1}(t)-(J(J+1)-M(M-1))P^{(\sigma)}_{J,M}(t)]\\&-\underbrace{2J}_{(A)}\gamma^\sigma_d\left[\underbrace{1}_{(B)}\underbrace{\left(1-\left|\frac{M}{J}\right|^2\right)}_{(C)}P^{(\sigma)}_{J,M}(t)\underbrace{+2}_{(D)}\left(J+\frac{1}{2}\right)\underbrace{\left(1-\left|\frac{M+\frac{1}{2}}{J+\frac{1}{2}}\right|^2\right)}_{(E)}P^{(\sigma)}_{J+\frac{1}{2},M+\frac{1}{2}}(t)\right]\\&+\gamma^\sigma_{ISC}\left[\underbrace{(J+M+1)}_{(F)}P^{(\sigma)}_{J+\frac{1}{2},M+\frac{1}{2}}(t)-\underbrace{(J+M)}_{(G)}P^{(\sigma)}_{J,M}(t)\right].\end{aligned}\tag{1}$$

In Eq. (1), we have highlighted specific terms and labeled them with consecutive letters using brackets. These terms are identified as flawed and necessitate modification. By marking them in this manner, we facilitate precise and convenient reference and commentary on these terms in subsequent sections of the article.

At this point, we explain all undefined notations appearing in the aforementioned equation and in all subsequent equations. To facilitate seamless comparison and ensure clarity, we adopt a convention identical to that of ref. 1:

- $P^{(\sigma)}_{J,M}$ denotes the population of the collective state with the numbers $J$ and $M$ in the collective space,
- $\gamma$ represents the decay constant,
- $\gamma^\sigma_{ISC}$ signifies the so-called intersystem crossing (ISC) rate,
- $\gamma^\sigma_d$ refers to the dephasing rate,
- $N^{(\sigma)}_{nc}$ stands for the number of independently radiating emitters (exponentially) at time t
- $F_{N_\sigma}$ represents the fluorescence at a given time t,
- $N$ signifies the total number of emitters.

The second important formula provides the number of non-radiative decaying emitters – denoted by $N_{nc}$ (Eq. 13 in ref. 1), which takes the form:

$$\frac{d}{dt}N^{(\sigma)}_{nc}=-(\gamma+\gamma^\sigma_{ISC})N^{(\sigma)}_{nc}(t)+\gamma^\sigma_d\sum_{J=\frac{1}{2}}^{\frac{N}{2}}\sum_{M=-J}^{J}\underbrace{\left(1-\left|\frac{M}{J}\right|^2\right)}_{(H)}2JP^{(\sigma)}_{J,M},\tag{2}$$

where a specific term has been highlighted for reference.

The last key equation describes the fluorescence rate (Eq. 15 in ref. 1) and takes the form:

$$F_{N_\sigma}(t)=\gamma(N^{(\sigma)}_{nc}(t)+\sum_{J=\frac{1}{2}}^{\frac{N}{2}}\sum_{M=-J}^{J}(J(J+1)-\underbrace{M(M+1)}_{(I)})P^{(\sigma)}_{J,M}).\tag{3}$$

These three equations are fundamental for the results presented in ref. 1. Eq. (1) describes the dynamics of the occupation of collective states $|J,M,\sigma\rangle$ with quantum numbers $J$, $M$ under the influence of mechanisms such as spontaneous emission, dephasing, or radiative transitions. Eq. (2) describes the population of emitters that radiate light according to the exponential decay law outlined in the Weiskopf-Wigner theory[7]. The effects of dephasing on the collective state of the color centers in the diamond lead to an increase in the number of such emitters, resulting in a decrease in the number of collectively emitting emitters. Eq. (3) describes the fluorescence of light based on predictions of the population of collectively and standardly emitting emitters. With such tools, the intensity of light can be plotted and observed.

## Model A: Critical assessment

In this paragraph, we present fluorescence plots generated from analytical formulas provided in ref. 1. For consistency, we employ identical parameters as those specified in ref. 1. By juxtaposing the obtained results with the data and simulations from ref. 1, we aim to visually demonstrate the inaccuracies of Model A. Given that a single clear discrepancy suffices, we show the case involving 7 color centers. We have also tested examples involving 2 and 10 centers, both of which exhibit inconsitencies. However, to keep the present section concise, these findings are presented in Supplementary Information.

In the scenario involving 7 color centers ($N$ = 7), we implemented Models A and B, and the corresponding results are depicted in Fig. 1a. In this plot, the solid line corresponds to Model A. One can notice that the observed fluorescence pattern does not align with the expected behavior of Dicke's superradiance, which is the focus of our discussion. Instead, we observe intense fluorescence from the initial moment, reaching a maximum value of 1 for $t$ = 0, and then declining rapidly. The plot generated from Model A differs significantly from the results provided in ref. 1 for the same parameters, cf. Fig. 1b on the right side. Notably, our simulations using Model B, represented by the dashed line in Fig. 1a on the left side, yielded contrasting plots compared to those obtained by Model A.

Additionally, we encountered instances of unphysical behavior in the simulation results generated by Model A. Specifically, we observed that the fluorescence intensity dropped below zero within a certain time interval, as indicated by the vertical lines in Fig. 1a. Furthermore, an unphysical non-zero asymptote was obtained. To further examine this phenomenon, we extended the time range, as illustrated in additional figures provided in Supplementary Information.

## Model B: Superadiance from a single diamond nanocrystal

According to our research, the proper master equation for fluorescence modeling reads:

$$\begin{aligned}\frac{d}{dt}P^{(\sigma)}_{J,M}(t)=&\,\gamma\Big[(J(J+1)-M(M+1))P^{(\sigma)}_{J,M+1}(t)-(J(J+1)-M(M-1))P^{(\sigma)}_{J,M}(t)\Big]\\&-\gamma^\sigma_d\left[2J\left|\frac{M}{J}\right|^2P^{(\sigma)}_{J,M}(t)-2\left(J+\frac{1}{2}\right)\left|\frac{M+\frac{1}{2}}{J+\frac{1}{2}}\right|^2P^{(\sigma)}_{J+\frac{1}{2},M+\frac{1}{2}}(t)\right]\\&+\gamma^\sigma_{ISC}\Big[(J+M+1)(J-M+1)P^{(\sigma)}_{J+\frac{1}{2},M+\frac{1}{2}}(t)-(J+M)(J-M+1)P^{(\sigma)}_{J,M}(t)\Big].\end{aligned}\tag{4}$$

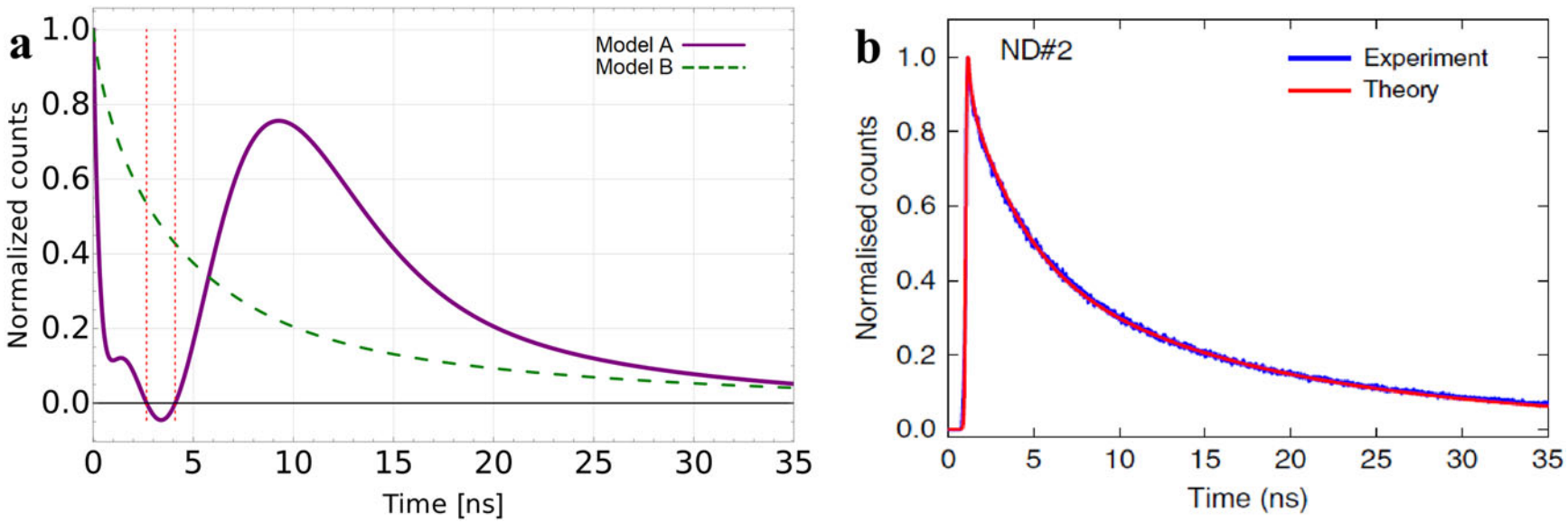

**Fig. 1 | Comparison of fluorescence models. a** Simulated fluorescence intensity over time, from 0 to 35 ns, for 7 NV centers. The purple solid line represents Model A[1], and the green dashed line represents Model B introduced in this paper. **b** Simulation results of Model A against experimental data, taken from ref. 1.

Of particular significance are the initial conditions we imposed on populating states within the collective space. We followed the same approach as the authors of the referenced work[1], assuming an even distribution of color centers across individual states. Consequently, the initial state of the total system of color centers is described as follows:

$$\widehat{\rho}^{(\sigma)}(0) = \sum_{N_\sigma} p_{N_\sigma} \widehat{\rho}^{(\sigma)}_{N_\sigma}(0), \tag{5}$$

where $p_{N_\sigma}$ represents the probability that a domain of color centers, characterized by its size $N_\sigma$, exists with the spin $\sigma$. The summation over these probabilities yields the probability of having a total spin $\sigma$, i.e. $\sum_{N_\sigma} p_{N_\sigma} = p_\sigma$. The properties of the density matrix $\widehat{\rho}^{(\sigma)}_{N_\sigma}(0)$ should be further explained. It can be decomposed as a statistical mixture over projectors on Dicke states $|J, M, \sigma\rangle\langle J, M, \sigma|$:

$$\widehat{\rho}^{(\sigma)}_{N_\sigma}(0) = \sum_M P^{(\sigma)}_{J=\frac{N_\sigma}{2}, M}(0) \left| \frac{N_\sigma}{2}, M, \sigma \right\rangle \left\langle \frac{N_\sigma}{2}, M, \sigma \right|, \tag{6}$$

where the occupation probability in subspaces, $P^{(\sigma)}_{J=\frac{N_\sigma}{2}, M}(0)$, is equal to $(N_\sigma + 1)^{-1}$. Additionally, we sum over the two possible values of spin projection, resulting in $\sum_{\sigma=0,1} p_\sigma = 1$. Following ref. 1, we treat cases with spin projection $\pm 1$ as indistinguishable.

The number of independently emitting color centers increases proportionally to the parameter $\gamma_d^\sigma$ responsible for dephasing, accompanied by the appropriate probability coefficients. We have provided a deeper discussion about these quantities in Supplementary Information. For now, let us introduce the differential equation governing the population decay of independently radiating color centers, which is known for its exponential nature

$$\frac{d}{dt} N^{(\sigma)}_{nc} = -(\gamma + \gamma^\sigma_{ISC}) N^{(\sigma)}_{nc}(t) + \gamma^\sigma_d \sum_{J=\frac{1}{2}}^{\frac{N}{2}} \sum_{M=-J}^{J} \left| \frac{M}{J} \right|^2 2J \, P^{(\sigma)}_{J,M}. \tag{7}$$

The most crucial quantity for fluorescence modeling, which is also an important component of ref. 1, is the total (from all domains) fluorescence intensity, which can be determined from the following equation:

$$F_{N_\sigma}(t) = \gamma \left( N^{(\sigma)}_{nc}(t) + \sum_{J=\frac{1}{2}}^{\frac{N}{2}} \sum_{M=-J}^{J} (J(J+1) - M(M-1)) P^{(\sigma)}_{J,M} \right). \tag{8}$$

The numerical algorithms need to be developed, a task to be executed in the subsequent software implementation. What remains to be described is the structure of the states comprising our physical system. The states are derived from ref. 1 and are implemented in our numerical simulations, as detailed in Supplementary Information.

A comprehensive presentation and comparison of the master equations from Models A and B are provided in detail in Supplementary Information. Additionally, solid arguments supporting the need for significant and conceptual modifications of the theoretical framework proposed in ref. 1 are presented.

## Discussion

In this paper, we have presented a detailed assessment of the theoretical model proposed in ref. 1, labeled as Model A, regarding room-temperature superradiance from NV color centers in diamonds. We presented correct equations, Model B, which when evaluated by numerical means provide physically meaningful results aligned with experimental data.

Our investigation revealed several key findings that challenge the applicability of Model A. Firstly, we observed significant discrepancies between the simulated fluorescence patterns obtained from Model A and those reported in ref. 1. These discrepancies were particularly pronounced in scenarios involving a small number of color centers, such as the case of $N = 7$ emitters, where our simulations yielded markedly different results from those published in ref. 1.

Furthermore, we observed simulation outcomes that appear unphysical based on Model A. Specifically, we observed cases where the normalized counts dropped below zero, and an unphysical non-zero asymptote was obtained. These findings indicate limitations in the ability of Model A to capture the physical behavior of the system under consideration.

Our analysis goes beyond demonstrating numerical discrepancies and unphysical behavior. We also provided the correct model that remains in agreement with physical assumptions related to fluorescence. By rigorously comparing our theory with the framework given in ref. 1, we aimed to provide a comprehensive analysis of the limitations of Model A.

It is important to note that while our analysis challenges the validity of Model A, it does not negate the significance of the experimental findings reported in ref. 1. Experimental results play a crucial role in advancing our understanding of complex physical phenomena, and the discrepancies identified in the theoretical framework demonstrate the need for further refinement and validation of modeling.

## Data availability

This manuscript has no associated data. The data statement is not applicable to this article as no datasets were generated or analyzed during the current study.

## Code availability

The numerical code used to generate the results presented in this study has been deposited as open-source under the license CC BY 4.0 and is available from ResearchGate: Borkowski, J. J., Czerwinski, A. & Kolenderski, P. Simulation Code for Superradiant Fluorescence from NV Centers in Diamonds. ResearchGate (2025). https://doi.org/10.13140/RG.2.2.12161.11364.

## Acknowledgements

This research was supported by the National Science Centre of Poland (Grant No. 2020/39/O/ST2/00137).

## Author contributions

Conceptualization: J.J.B. and P.K. Investigation: J.J.B., A.C., and P.K. Code development: J.J.B. Validation: A.C. and P.K. Supervision: A.C. and P.K. Funding acquisition: P.K. Writing - original draft: J.J.B., A.C., and P.K. Writing - review & editing: A.C.

## Competing interests
The authors declare no competing interests.

## Additional information
**Supplementary information** The online version contains supplementary material available at https://doi.org/10.1038/s41467-025-67847-6.

**Correspondence** and requests for materials should be addressed to Jakub J. Borkowski, Artur Czerwinski or Piotr Kolenderski.

**Peer review information** *Nature Communications* thanks Akbar Safari and the other anonymous reviewer for their contribution to the peer review of this work.